\documentclass[runningheads,a4paper]{llncs}

\usepackage{amssymb,amsmath}
\usepackage{graphicx,epsfig}

\newcommand{\keywords}[1]{\par\addvspace\baselineskip
\noindent\keywordname\enspace\ignorespaces#1}

\newcommand{\beq}{\begin{equation}}
\newcommand{\bra}[1]{\ensuremath{\langle\, #1\,\vert}}
\newcommand{\braket}[2]{\ensuremath{\langle\,#1\,\vert\,#2\,\rangle}}
\newcommand{\dee}[2]{\ensuremath{{{\rm d} #1\over {\rm d} #2}}}
\newcommand{\doo}[2]{\ensuremath{{\partial #1\over \partial #2}}}
\newcommand{\eeq}{\end{equation}}
\newcommand{\fundoo}[2]{\ensuremath{{\delta #1\over \delta #2}}}
\newcommand{\ket}[1]{\ensuremath{\vert\,#1\,\rangle}}
\newcommand{\matrel}[3]{\ensuremath{\left\langle\,#1\,\left\vert\,#2\,\right\vert\,#3\,\right\rangle}}
\newcommand{\Tr}{\mathrm{Tr\,}}

\def\ann{{\hat{a}}}
\def\cre{{\hat{a}^+}}
\def\Lio{{\hat{L}}}
\def\oA{\hat{A}}
\def\ophi{\hat{\varphi}}
\def\oPhi{\hat{\phi}}
\def\opi{\hat{\pi}}
\def\orho{\hat{\rho}}
\def\oU{\hat{U}}
\def\vx{{\bf x}}
\def\vy{{\bf y}}

\begin{document}

\mainmatter  

\title{Functional Methods in Stochastic Systems}

\titlerunning{Functional Methods in Stochastic Systems}

%
\author{Juha Honkonen}%
\authorrunning{Juha Honkonen}
%

\institute{National Defence University\\
P.O. Box 7, 00861 Helsinki, Finland}

%
%

\toctitle{Lecture Notes in Computer Science}
\tocauthor{Authors' Instructions}
\maketitle

\begin{abstract}
Field-theoretic construction of functional representations of
solutions of stochastic differential equations and master equations
is reviewed. A generic expression for the generating
function of Green functions of stochastic systems is put forward.
Relation of ambiguities in stochastic differential
equations and in the functional representations is discussed.
Ordinary differential equations for expectation values and
correlation functions are inferred with the aid of a variational
approach.
\end{abstract}

\section{Introduction}
\label{se:intro}

Fluctuation effects in physics, chemistry, biology, operations research etc. are often described by the Langevin equation
(stochastic differential equation, SDE)
\beq
\label{Langevin0}
\doo{\varphi}{t}=-K\varphi+U(\varphi)+fb(\varphi)\,,
\eeq
where $f$ is a Gaussian random field with zero mean and the
white-in-time correlation function
\beq
\label{correlator}
\langle f(t,{\bf x})f(t',{\bf
x}')\rangle=\delta(t-t')D({\bf x}-{\bf x}')\,.
\eeq
The Langevin equation with
white noise is mathematically ill-defined. A consistent description may be obtained by using, e.g.,
a $\delta$ sequence with finite correlation times
\beq
\langle f(t,{\bf x})f(t',{\bf x}')\rangle=
\overline{D}(t,{\bf x};t',{\bf x}')\longrightarrow
\delta(t-t')D({\bf x},{\bf x}')\,,\nonumber
\eeq
in the construction of the expectation values of functions of $\varphi$ and passing to the white-noise limit in the end.
This is often physically natural and leads to solution, which corresponds to the Stratonovich interpretation of the
SDE. However, the point of introducing of the SDE with white noise is to avoid dealing with this limit
explicitly. Technically, the Stratonovich interpretation is complicated and the SDE is most often used
in the Ito interpretation in mathematical treatments.

Instead of using the mathematically problematic, although physically
transparent, Langevin equation the stochastic problem (\ref{Langevin0}), (\ref{correlator}) may be
equivalently stated in terms of the Fokker-Planck equation (FPE), which is an equation for
both the conditional probability density
$
p\left(\varphi,t\vert\varphi_0,t_0\right)
$
and the probability density
$
p\left(\varphi,t\right)
$
of the variable $\varphi$.
The simple way to demonstrate this equivalence uses rules of Ito calculus \cite{Gardiner97}, however, and
I am not going to dwell on this issue here. Other seemingly simple methods use -- at least implicitly --
functional integrals, which are still mathematically ill-defined, let alone the Gaussian integral.
Therefore, only the correspondence between the quantities
specifying the stochastic problem in both approaches will be quoted here. The main advantage of the
Fokker-Planck equation is that the equation itself is completely well-defined partial differential
(or functional-differential for field variables) equation. The ambiguity of the Langevin problem
shows in that the FPE is different for different interpretations of the SDE.

For simplicity of notation, consider zero-dimensional field theory.
The Fokker-Planck equation for the conditional probability density
$
p\left(\varphi,t\vert\varphi_0,t_0\right)
$
in the case of the Ito equation is
\begin{multline}
\label{FokkerPlanckIto0}
\doo{}{t}p\left(\varphi,t\vert\varphi_0,t_0\right)
=-\doo{}{\varphi}\left\{\left[-K\varphi+U(\varphi)\right]p\left(\varphi,t\vert\varphi_0,t_0\right)\right\}\\
+{1\over 2}\doo{^2}{\varphi^2}\left[b(\varphi)Db(\varphi)p\left(\varphi,t\vert\varphi_0,t_0\right)\right]\,.
\end{multline}
If the SDE (\ref{Langevin0}) is interpreted in the Stratonovich sense, the FPE is
\begin{multline}
\label{FokkerPlanckStrato0}
\doo{}{t}p\left(\varphi,t\vert\varphi_0,t_0\right)
=-\doo{}{\varphi}\left\{\left[-K\varphi+U(\varphi)\right]p\left(\varphi,t\vert\varphi_0,t_0\right)\right\}\\
+{1\over 2}\doo{}{\varphi}\left\{b(\varphi)\doo{}{\varphi}\left[Db(\varphi)p\left(\varphi,t\vert\varphi_0,t_0\right)\right]\right\}\,.
\end{multline}
In what follows I will use the fact that
the conditional probability density
$
p\left(\varphi,t\vert\varphi_0,t_0\right)
$
is the fundamental solution of the FPE (\ref{FokkerPlanckIto0}) or (\ref{FokkerPlanckStrato0}), i.e.
\beq
p\left(\varphi,t_0\vert\varphi_0,t_0\right)=\delta\left(\varphi-\varphi_0\right)\,.\nonumber
\eeq
Contractions are not quite obvious, when the random variable has several components. For instance, the Fokker-Planck equation in the Ito form
becomes
\begin{multline}
\doo{}{t}p\left(\varphi,t\vert\varphi_0,t_0\right)
=-\doo{}{\varphi_i}\left\{\left[-K_{ij}\varphi_j+U_i(\varphi)\right]p\left(\varphi,t\vert\varphi_0,t_0\right)\right\}\\
+{1\over 2}\doo{^2}{\varphi_i\partial\varphi_j}\left[b_{ik}(\varphi)D_{kl}b_{jl}(\varphi)p\left(\varphi,t\vert\varphi_0,t_0\right)\right]\nonumber
\end{multline}
for a multi-component variable $\varphi_i$.

The Fokker-Planck equation is similar to Schr\"odinger equation. Using this analogy, the solution of the
FPE as well as calculation of expectation values may be represented in a way analogous to quantum
mechanics \cite{Leschke77}. Construction with the FPE as the starting
point gives rise to the famous Martin-Siggia-Rose solution of the SDE \cite{MSR}, but avoids ambiguities
inherent in the SDE.

Markov processes described in terms of the Fokker-Planck equation
have continuous sample paths, i.e. $\varphi=\varphi(t)$ is a continuous function
of time. Not all interesting stochastic processes
belong to this category. A wide class of such processes describe changes
in occupation numbers (e.g. individuals of some population, molecules in chemical reaction)
which cannot be naturally described by continuous paths.
This kind of processes are described by {\em master equations} -- a special case of
(differential) Kolmogorov equations \cite{Gardiner97}.

The generic form of a master equation
written for the conditional probability density $p\left(\varphi,t\vert\varphi_0,t_0\right)$ of
a Markov process is
\begin{equation}
\doo{}{t}p\left(\varphi,t\vert\varphi_0,t_0\right)
=\int\!d\chi\left[W(\varphi\vert \chi,t)p\left(\chi,t\vert\varphi_0,t_0\right)-
W(\chi\vert\varphi ,t)p\left(\varphi,t\vert\varphi_0,t_0\right)\right]\,,\nonumber
\end{equation}
where
$W(\varphi\vert \chi,t)$
is the {\em transition probability} per unit time, whose formal definition from
the differential Kolmogorov equation is (for all $\varepsilon>0$)
\[
W(\varphi\vert \chi,t)=\lim_{\Delta t\to 0}{p\left(\varphi,t+\Delta t\vert\chi,t\right)\over \Delta t}\,,
\]
uniformly in $\varphi$, $\chi$ and $t$ for all $\vert \varphi-\chi\vert \ge\varepsilon$.

I shall be using the master equation for discrete variables (occupation numbers). In this case the discontinuous
character of the paths of the {\em jump processes} described by the master equation is especially transparent.
The transition probabilities are usually simple functions of the occupation number $n$.
As an example, consider the generic master equation for the
{\em stochastic Verhulst model}
\begin{multline}
\label{VerhulstGen}
\dee{P(t,n)}{t}=[\beta
(n+1)+\gamma(n+1)^2]P(t,n+1)+\lambda(n-1)P(t,n-1)\\
 -\left(\beta
n+\lambda n+ \gamma n^2\right)P(t,n)\,,
\end{multline}
where $\beta$ is the death rate, $\lambda$ the birth rate and
$\gamma$ the damping coefficient necessary to bring about a saturation for
the population.
The choice $\lambda=\beta=0$ leads to the master equation for
the annihilation reaction $A+A\to A$.
The set of master equations may also be cast in the form of an evolution equation of the type of
Schr\"odinger equation in a Fock space of many-particle quantum mechanics. This representation is due
to Doi \cite{Doi76}. Contrary to the widely known Martin-Siggia-Rose approach, the method of Doi has
been gaining due attention only recently.

The two quantum-mechanical treatments of stochastic problems have rather different appearances in the
literature. The first aim of this paper is to present both cases in a unified fashion in terms of generating
functions of Green functions of many-body quantum mechanics. Second, ambiguities in the functional representation
of the solution of the stochastic problems are expressed in terms of the thoroughly analyzed functional form of
the quantum field theory. Third,
the functional representation is used
to calculate fluctuation corrections to rate equations.

This paper is organized as follows. In Section \ref{sec:OperatorForm}
the operator representation of stochastic problems and their connection to Green functions is reviewed.
Section \ref{sec:Greens} is devoted to discussion of functional representations of Green functions.
Variational method in the functional approach is discussed in Section \ref{sec:FlucAmendedRates} and
conclusions presented in Section \ref{sec:Conclusion}.

\section{Quantum-Mechanical Representation of Stochastic Problems}
\label{sec:OperatorForm}

\subsection{Green Functions for the Fokker-Planck Equation}

Consider, for definiteness, the Fokker-Planck equation (\ref{FokkerPlanckIto0}) corresponding to the Ito
interpretation of the Langevin equation (\ref{Langevin0}).
Introduce -- in analogy with Dirac's notation in quantum mechanics -- the state vector $\ket{p_t}$ according to
the following representation of the PDF
\[
p(\varphi,t)=\braket{\varphi}{p_t}\,,
\]
which is the solution of the FPE (\ref{FokkerPlanckIto0}) with the initial condition $p(\varphi,0)=p_0(\varphi)$.
To construct the evolution operator for the state vector, introduce momentum and coordinate operators
in the manner of quantum mechanics by relations
\[
\hat{\pi}f(\varphi)=-\doo{}{\varphi}f(\varphi)\,,\quad \hat{\varphi}f(\varphi)=\varphi f(\varphi)\,,\quad \left[\hat{\varphi},\hat{\pi}\right]=1\,.
\]
In these terms, the FPE for the PDF gives rise to the evolution equation for the state vector in the form
\[
\doo{}{t}\ket{p_t}=\hat{L}\ket{p_t}\,,
\]
where the Liouville operator for the FPE corresponding to the Ito interpretation of the SDE assumes, according to (\ref{FokkerPlanckIto0}), the form
\beq
\label{LioFPEIto}
\Lio=
\opi\left[-K\ophi+U(\ophi)\right]
+{1\over
2}\opi^2b(\ophi)Db(\ophi)\,.
\eeq
Note that, contrary to quantum mechanics, there is no ordering ambiguity in the construction
of the Liouville operator here.

In this notation, the conditional PDF may be expressed as the matrix element
\beq
\label{CondPDFMatrEl}
p\left(\varphi,t\vert\varphi_0,t_0\right)=\matrel{\varphi}{e^{\Lio(t-t_0)}}{\varphi_0}\,.
\eeq
Introduce time-dependent operators $\hat{\varphi}(t)$ in the Heisenberg picture
of imaginary time quantum mechanics (i.e. Euclidean quantum mechanics):
\beq
\label{LioOpe}
\hat{\varphi}_H(t)= e^{-\hat{L}(t-t_0)}\hat{\varphi}e^{\hat{L}(t-t_0)}\,,
\eeq
and define the time-ordered product (chronological product, $T$ product) of time-dependent operators
\beq
\label{TprodLio}
T\left[\oA_1(t_1)\cdots\oA_n(t_n)\right]=\sum\limits_{P(1,\ldots,n)}
P\left[\theta\left(t_1\ldots t_n\right)\oA_1(t_1)\cdots\oA_n(t_n)\right]\,,
\eeq
where
\[
\theta\left(t_1\ldots t_n\right)\equiv\theta\left(t_1-t_2\right)\theta\left(t_2-t_3\right)\cdots
\theta\left(t_{n-1}-t_n\right)\,.
\]
In definition (\ref{TprodLio}) the sum is taken over all permutations of the labels of
the time arguments $\left\{t_i\right\}_{i=1}^n$ and the operators in each term are put in the
order of growing time arguments from the right to the left. Thus, under the $T$-product sign operators
commute. It should be noted that the definition of the time-ordered product
should be amended for coinciding time arguments.

Introduce then the $n$-point Green function
as the quantum-mechanical expectation value of the $T$-product of $n$ operators (\ref{LioOpe})
\beq
\label{GnLio}
G_n(t_1,t_2,\ldots t_n)=\Tr\left\{\hat{p}_0T\left[\ophi_H(t_1)\ophi_H(t_2)\cdots\ophi_H(t_n)\right]\right\}
\eeq
determined by the trace $\Tr$ and the density operator
\beq
\label{IniDensityOpe}
\hat{p}_0=\int\!d\varphi\ket{p_0}\bra{\varphi}\,.
\eeq
Choosing, for definiteness, the time sequence $t_1> t_2> t_3> \ldots > t_{n-1}> t_n>t_0$ it is readily seen by
direct substitution of relations (\ref{CondPDFMatrEl}), (\ref{LioOpe}) and (\ref{IniDensityOpe}) in (\ref{GnLio})
with the aid of the
normalization conditions of the PDF
and insertions of the resolution of the identity
$
\int\!d\varphi \,\ket{\varphi}\bra{\varphi}=1
$
that
\beq
\label{MomentGF}
\int\!d\varphi_1\ldots\int\!d\varphi_n\,\varphi_1\cdots\varphi_n
p\left(\varphi_1,t_1;\varphi_2,t_2;\ldots ;\varphi_n,t_n\right)=G_n(t_1,t_2,\ldots t_n)\,,
\eeq
i.e. the Green function (\ref{GnLio}) is equal to the moment function (\ref{MomentGF}). This relation connects the
operator approach to evaluation of moments of the random process.

\subsection{Green Functions for the Master Equation}

The set of master equations for $P(t,n)$ is reduced to a single equation by ''second quantization'' of Doi.
Build a Fock space: operators $\ann$, $\cre$ and basis vectors $\ket{n}$:
\beq
\ann\ket{0}=0\,,\quad \cre\ket{n}=\ket{n+1}\,,\quad
[\,\ann,\cre]=1\,,\quad \braket{n}{m}=n!\delta_{nm}\,.\nonumber
\eeq
Master equations yield kinetic equation
for the state vector
\begin{equation}
\label{statevector}
\ket{P_t}=\sum\limits_{n=0}^\infty P(t,n)\ket{n}\,.
\end{equation}
in the form of a single evolution equation for the state vector (\ref{statevector})
without any explicit dependence on the occupation number:
\begin{equation}
\label{L}
\dee{\ket{P_t}}{t}=\Lio(\cre,\,\ann)\ket{P_t}\,,
\end{equation}
where the Liouville operator $\Lio(\cre,\,\ann)$
is constructed from the set of master equations according to rules:
\begin{align*}
nP(t,n)\ket{n}&=\cre\ann P(t,n)\ket{n}\,,\\
{n}P(t,n){\ket{n-1}}&=
{\ann} P(t,n){\ket{n}}\,,\\
nP(t,n){\ket{n+1}}&=
{\cre}\cre\ann P(t,n){\ket{n}}\,.
\end{align*}
For instance, the Liouville operator for the stochastic Verhulst model (\ref{VerhulstGen}) is
\begin{equation}
\label{VerhulstLio}
\Lio(\cre,\ann)= \beta(I-\cre)\ann
+\gamma(I-\cre)\ann\cre\ann +\lambda(\cre-I)\cre\ann\,.
\end{equation}
Equation (\ref{L}) gives rise to the Heisenberg
evolution of operators in analogy with (\ref{LioOpe}).
To represent expectation values of occupation-number dependent quantities
in the operator formalism use the projection vector $\bra{P}$:
\begin{equation}
\label{Pvec}
\bra{P}=\sum\limits_{n=0}^\infty{1\over n!}\,\bra{n}=\sum\limits_{n=0}^\infty{1\over n!}\,\bra{0}\ann^n
=\bra{0}e^\ann\,.
\end{equation}
Consider the Green function of occupation-number operators $\hat{n}_H(t)=\cre_H(t)\ann_H(t)$:
\beq
\label{GnLioMaster}
G_m(t_1,t_2,\ldots t_m)=\Tr\left\{\hat{P}_0T\left[\hat{n}_H(t_1)\hat{n}_H(t_2)\cdots\hat{n}_H(t_m)\right]\right\}\,,
\eeq
with the density operator
\beq
\label{IniDensityOpeMaster}
\hat{P}_0=\ket{P_0}\bra{P}\,.
\eeq
From definitions it follows that the conditional probability density function for the master equation may be written
as (the factorial in front of the matrix element is due to the unusual normalization of the basis states)
\beq
\label{CondPDFMatrElMaster}
P\left(n,t\vert n_0,t_0\right)={1\over n!}\,\matrel{n}{e^{\Lio(t-t_0)}}{n_0}\,.
\eeq
Choosing, for definiteness, the time sequence $t_1> t_2> t_3> \ldots > t_{n-1}> t_n>t_0$ it is readily seen by
direct substitution of relations (\ref{CondPDFMatrElMaster}) and (\ref{IniDensityOpeMaster}) in (\ref{GnLioMaster})
with the aid of the
normalization conditions of the PDF
and insertions of the resolution of the identity
$
\displaystyle \sum_n {1\over n!}\,\ket{n}\bra{n}=1
$
that
\beq
\label{MomentGFMaster}
\sum_{n_1}\ldots\,\sum_{n_m}\,{n_1}\cdots n_m
P\left(n_1,t_1;n_2,t_2;\ldots ;n_m,t_m\right)=G_m(t_1,t_2,\ldots t_m)\,,
\eeq
i.e. the Green function (\ref{GnLioMaster}) is equal to the moment function (\ref{MomentGFMaster}). This relation connects the
operator approach to evaluation of moments of the random process described by a master equation.

\section{Functional Representatation of Green Functions}
\label{sec:Greens}

The generic representation for the generating function of Green functions introduced in the previous section may be
written in the form
\beq
\label{GenericG}
G(J)=\Tr\left[\hat{\rho}_0T\,e^{\hat{S}_J}\right]\,,
\eeq
where the source term and the density operator are
\beq
\hat{S}_J=\int_{t_i}^{t_f}\!dt\,\ophi_H(t) J(t)\,, \qquad \orho_0=
\int\!d\varphi\ket{p_0}\bra{\varphi}\nonumber
\eeq
for the Fokker-Planck equation,
or, for the master equation,
\beq
\hat{S}_J=\int_{t_i}^{t_f}\!dt\,\cre_H(t)\ann_H(t) J(t)\,,\qquad
\orho_0=\ket{P_0}\bra{P}\,.\nonumber
\eeq
Perturbation theory is constructed in the Dirac (interaction) picture of quantum mechanics.
To this end, the Liouville operator is decomposed to the free and interaction parts
$\Lio_0$ and
$\Lio_I$:
\begin{equation}
\Lio=\Lio_0+\Lio_I\,.\nonumber
\end{equation}
In our cases, the convenient choices are $\Lio_0=-K\opi\ophi$ for (\ref{LioFPEIto})
and $\Lio_0=-\beta\cre\ann$ for (\ref{VerhulstLio}), where $K>0$, $\beta>0$.
Time-dependent operators in the interaction picture are defined according to
\begin{equation}
\ophi(t)=e^{-(t-t_0)\Lio_0}\ophi e^{(t-t_0)\Lio_0}\,.\nonumber
\end{equation}
The corresponding evolution operator may be expressed in terms of the $T$-product (here, $t>t'>t_0$):
\begin{multline}
\label{evoFPTexp}
\oU ( t-t_0 , t '-t_0)=
e^{-(t-t_0)\Lio_0 }e^{( t - t ')\Lio}e^{(t'-t_0)\Lio_0}\\
=e^{\hat{L}_0t_0}T\exp\left[\,\int_{ t '}^{ t}
\!\Lio_I(u)\,du\right]e^{-\hat{L}_0t_0}\,.
\end{multline}
Note that the Dirac operators in the
$T$-product do not carry any $t_0$-dependence in (\ref{evoFPTexp}).
In the interaction picture the  $T$-product in (\ref{GenericG}) may be written as \cite{Vasilev98}
\beq
\label{TtoT}
T\,e^{\hat{S}_J}
=e^{\hat{L}_0t_0}\oU(t_0,t_f)
T\left[e^{\hat{S}_J+\hat{S}_I}
\right]\oU(t_i,t_0)e^{-\hat{L}_0t_0}\,,
\eeq
where
$\hat{S}_I=\int_{ t_i}^{t_f}\!\Lio_I(t)\,dt$
and the time instants $t_f$ and $t_i$ are chosen such that $t_f> t_l>t_i>t_0$ for all $t_l$, $l=1,2,\ldots ,n$.
The evolution operator with the reversed order of time arguments may be cast into the form
of the
{\em anti-chronologically ordered}$\,$ exponential ($t>t_0$)
\beq
\label{antiT}
\oU(t_0,t)=\tilde{T}\,e^{-\int_{t_0}^t\!\Lio(t)\,dt}\,.
\eeq
With the use of representations (\ref{evoFPTexp}) and (\ref{antiT}) the chronological product
in the generating function (\ref{GenericG}) may be expressed as the product of three chronological
products:
\beq
\label{TtoT3}
T\,e^{\hat{S}_J}
=e^{\hat{L}_0t_0}
\tilde{T}\,e^{-\int_{t_0}^{t_f}\!\Lio(t)\,dt}
T\left[e^{\hat{S}_J+\hat{S}_I}\right]
{T}\,e^{\int_{t_0}^{t_i}\!\Lio(t)\,dt}
e^{-\hat{L}_0t_0}\,,
\eeq
The three $T$-products in (\ref{TtoT3})
fuse -- due to Wick's theorems -- in a {\em normal product} ($\opi$ to the left of $\ophi$ or
$\cre$ to the left of $\ann$)  giving
rise to the representation \cite{Vasilev98}
\begin{multline}
\label{GtoN}
G(J)=\Tr e^{-\hat{L}_0t_0}\hat{\rho}_0e^{\hat{L}_0t_0}
N\biggl\{
 \exp\left[{1\over 2}\,\fundoo{}{\phi_1}\tilde{\Delta}\fundoo{}{\phi_1}
+{1\over 2}\,\fundoo{}{\phi_2}{\Delta}\fundoo{}{\phi_2}
+\fundoo{}{\phi_1}n\fundoo{}{\phi_2}
\right]
 \\ \times
 \exp\biggl[S_J(\phi_2)
 -\int_{t_0}^{t_f}\!{L_I}(\phi_1)\,du
 +\int_{t_ 0}^{t_f}\!L_I(\phi_2)\,du\biggr]
\biggr\vert_{\phi_1=\phi_2=\hat{\phi}}\biggr\}\,,
\end{multline}
where
$\hat{\phi}$ is a two-component shorthand for all the operators  appearing in $\Lio_I$, i.e.
either $\hat{\phi}=(\ophi,\opi)$ or  $\hat{\phi}=(\ann,\cre)$. In (\ref{GtoN}), the auxiliary field
variables $\phi_1$ and $\phi_2$ correspond to functional arguments prescribed to the anti-chronological
product and the chronological products in (\ref{TtoT3}), respectively. Originally, in the functional representation of
Wick's theorem for (\ref{TtoT3}) to each $T$-product a separate field variable is prescribed, but
those corresponding to the two consecutive rightmost $T$-product factors may be replaced by a single variable $\phi_2$.
The propagators and contractions in (\ref{GtoN}) are standard:
\begin{align*}
\tilde{\Delta}(t,t')&=\tilde{T}\left[\oPhi(t)\oPhi(t')\right]-N\left[\oPhi(t)\oPhi(t')\right]\,,\\
{\Delta}(t,t')&={T}\left[\oPhi(t)\oPhi(t')\right]-N\left[\oPhi(t)\oPhi(t')\right]\,,\\
n(t,t')&=\oPhi(t)\oPhi(t')-N\left[\oPhi(t)\oPhi(t')\right]\,.
\end{align*}
The functional $L_I$ in (\ref{GtoN}) (not an operator any more) is ambiguous, however, because the
chronological products have not been defined at coinciding time arguments. The choice of the
value of the $T$ and $\tilde{T}$ products at coinciding time arguments
fixes the form of the interaction functional $L_I$. It should be noted that this choice affects the
propagators as well. The {\em normal form}
obtained by replacing operators by functions in the representation obeying
$\Lio_I=N[\Lio_I]$ (see \cite{Vasilev98} for details) is the most natural here.
For the $T$-product this boils down to choosing the temporal step function
equal to zero at the origin. Thus, the practical rule to resolve this ambiguity sounds similar to that in the Ito interpretation
of the SDE \cite{Gardiner97}, but it should be borne in mind that that the latter ambiguity has already been fixed by different means and
there is no obligation to make the same choice here. Henceforth, the normal form of $L_I$ is implied together with the
corresponding choice: ${\Delta}(t,t)=0$, $\tilde{\Delta}(t,t)=0$.

To cast the generating function in the functional form without operators, the expression
\beq
\label{residue}
\Tr e^{-\hat{L}_0t_0}\hat{\rho}_0e^{\hat{L}_0t_0}N[\ldots]
\eeq
should be calculated. Here, the ellipsis under the normal product sign denotes an arbitrary operator.
This calculation is different in the two cases discussed. Moreover, in practical
calculations for the master equation it is convenient to pull the coherent state exponential of
the projection vector (\ref{Pvec}) through all the operator expressions to the rightmost ket-vector
of the state vector (\ref{statevector}). This leads to the shift of all creation operators
by the unit operator: $\cre\to \cre+1$. I shall keep the free Liouville operator in the form
chosen above and put all changes in the interaction operator: $\Lio\to \Lio_0+\Lio'_I$.
After these manipulations the density operator in (\ref{residue}) for the master equation
case assumes the form (strictly speaking, this is not a density operator any more)
\beq
\orho_0=\ket{P'_0}\bra{0}=\sum_{n=0}^\infty P(0,n)\left[\cre+1\right]^n\ket{0}\bra{0}\,.\nonumber
\eeq
It is an important property of the field theory of these stochastic problems that in representation (\ref{GtoN})
the terms corresponding to the chronological product (expressed in terms of the field $\phi_2$)
and the anti-chronological product ($\phi_1$) effectively factorize and the latter gives rise to the unity.
This is different from the Green functions of quantum kinetic theory based on a similar expression.
The factorization of the anti-chronological part in (\ref{GtoN}) is a consequence of three features. First,
only contractions $n(t,t')=\ophi(t)\opi(t')-N\left[\ophi(t)\opi(t')\right]$ and
$n(t,t')=\ann(t)\cre(t')-N\left[\ann(t)\cre(t')\right]$ are non-trivial, all the rest vanish; second, the average (\ref{residue}) does not add anything
to propagators and, third, in the interaction terms the two fields $\phi_1$, $\phi_2$ are not mixed.

Results of calculation of (\ref{GtoN}) will be quoted here for two important cases: First, the probability
distribution is assumed to be homogenous, i.e. that there exists an equilibrium limit
\beq
p\left(\varphi,t\vert\varphi_0,t_0\right)\to p_e(\varphi)\,, \quad t_0\to -\infty\,,\nonumber
\eeq
second, the limit $t_0,t_i\to -\infty$, which allows to consider the problem on the whole time axis.

With the use of the standard manipulations \cite{Vasilev98} in the calculation of (\ref{residue}) I obtain
in the case of the Fokker-Planck problem the representation
(with the choice $t_f\to\infty$)
\begin{multline}
\label{GnLio5}
G(J)
=\int\!d\eta\,p_e(\eta)\,
\exp\left[\,\fundoo{}{\varphi}\Delta
 \fundoo{}{\pi}\right]\\
 \times
\exp \left\{\,\int_{t_i}^{\infty}\left[
L_I(t) +\varphi(t) J(t)\right]\,dt\right\}\biggl\vert_{\varphi=\varphi_\eta(t-t_i)\atop
\pi=0\phantom{\varphi_\eta(t-t_i} }\,,
\end{multline}
where $\varphi_\eta$ is the solution of the equation $(\partial_t+K)\varphi_\eta(t)=0$ with the
initial condition $\varphi_\eta(0)=\eta$. Here, the choice $t_i=0$ gives rise to the standard Cauchy problem,
whereas in the limit $t_i\to -\infty$ leads to the ''scattering'' problem on the whole time axis
with the limit $\varphi_\eta\to 0$. In that case the normalization condition of $p_e(\eta)$ yields
\beq
\label{GS}
G(J)
=
\exp\left[\,\fundoo{}{\varphi}\Delta
 \fundoo{}{\pi}\right]\,
\exp \left\{\,\int_{-\infty}^{\infty}\left[
 L_I(t) +\varphi(t) J(t)\right]\,dt\right\}\biggl\vert_{\varphi=0\atop
\pi=0}\,,
\eeq
which, in fact, is the standard functional representation of quantum field theory (QFT) \cite{Vasilev98} and allows to use the methods
of the latter in practical calculations.

In the case of the master equation problem the corresponding functionals for the
generating function are
\begin{multline}
\label{GMECauchy}
G(J)
=\int\!{ds\over 2\pi i}\,e^s\tilde{G}(s)\,
\exp\left[\,\fundoo{}{a}\Delta
 \fundoo{}{a^+}\right]\,\\
  \times
\exp \left\{\,\int_{t_i}^{\infty}\left[
L'_I(t) +\left[(a^+(t)+1)a(t)\right] J(t)\right]\,dt\right\}\biggl\vert_{a=a_s(t-t_i)\atop
a^+=0\phantom{a_s(t-} }\,,
\end{multline}
where
\[
\tilde{G}(s)=\sum_{n=1}^\infty{\Gamma(n)\over s^n}\,P(0,n-1)\,.
\]
and $a_s$ is the solution of the equation $(\partial_t+\beta)a_s(t)=0$ with the
initial condition $a_s(0)=s$. The generating function for the problem on the whole time
axis assumes the form
\begin{multline}
\label{GMES}
G(J)
=
\exp\left[\,\fundoo{}{a}\Delta
 \fundoo{}{a^+}\right]\,\\
  \times
\exp \left\{\,\int_{-\infty}^{\infty}\left[
L'_I(t) +\left[(a^+(t)+1)a(t)\right] J(t)\right]\,dt\right\}\biggl\vert_{a=0\atop
a^+=0}\,.
\end{multline}
Although these expressions are written for the zero-dimensional field theory, their
generalization to fields in finite-dimensional space is straightforward. In that case
$K$ and $\beta$ stand for positive-definite second-order differential operators.

Functional-integral representations for generating functions may be obtained with the aid of the Gaussian integral representation
\beq
\label{GaussianReduction}
\exp\left[\,\fundoo{}{\varphi}\Delta
 \fundoo{}{\pi}\right]=\det \Delta\iint\!{\mathcal D}a^+{\mathcal
D}a\,\,\exp\left[-a^+\Delta^{-1}a+a^+\fundoo{}{\pi}+a\fundoo{}{\varphi}\right]\,.
\eeq
of the functional differential operators (for brevity, only the case of the Fokker-Planck equation will be discussed here).
It should be noted that the kernel $\Delta^{-1}$ on the right side of (\ref{GaussianReduction})
contains a differential operator. Therefore,
the functional integral generated with the aid of (\ref{GaussianReduction}) must imply boundary
conditions in the space of integration which single out
the proper Green function of this differential operator as the propagator $\Delta$ on the left side \cite{Vasilev98}.

In addition to this, the choice of the value of the chronological product at coinciding time arguments
affects both the form of the interaction functional $\int L_I(t)dt$
and the value of the propagator $\Delta$ at coinciding time arguments.
As discussed above, this effect is explicit in the interaction
functional, although often overlooked.
It is not clear, however, how this
choice may be taken into account in the Gaussian integral (\ref{GaussianReduction}).
The ambiguity in the definition of the chronological product
bears resemblance with the Ito-Stratonovich ambiguity of the original Langevin equation,
because its solution boils down to the choice of the value of the step function at the origin. It should be
borne in mind, however, that I have got rid of the latter at the outset by starting from the Fokker-Planck equation.
It is worth recalling that at any rate the
perturbation expansion is independent of the choice of the amendment of the definition of the chronological product
and this feature must be reproduced by any alternative calculation of the functional integral for generating functions.

However, in the functional-integral representations emerging from (\ref{GnLio5}), (\ref{GS}) and (\ref{GaussianReduction})
ambiguities inherent in the functional solution of the stochastic problem are explicit only in the interaction functional.
Therefore, care has to be exercised to take
into account subtleties discussed above, when the functional integral is used for numerical calculation of the Green functions -- or for any
calculation other than the perturbation expansion. The main point of the formulae (\ref{GnLio5}) - (\ref{GMES}) is to
give a unified, compact and unambiguous representation of the perturbative solution of the stochastic problems discussed.

\section{Variational method}
\label{sec:FlucAmendedRates}

From the wealth of calculational methods of the QFT useful in stochastic problems \cite{Vasilev04} I would like to emphasize here the
variational approach.
The more or less standard steepest descent method has been applied to calculation of the functional integrals for generating
functions, e.g., in order to evaluate the asymptotic behaviour
of high orders of the perturbation expansion. Since the method is based on the knowledge of an explicit solution
of a nonlinear differential equation, useful results have been scarce. Apart form the zero-dimensional field theory,
only few cases -- based on the previously known instanton of the static $\varphi^4$ model --
have been analyzed for various models of critical dynamics \cite{Honkonen05a,Honkonen05b}.

The variational method may also be applied to construct fluctuation-amended rate equations and the like. To this end, the generating functional
$G$ is not suitable because it does not give rise to solutions in terms of moments directly. Such solutions may be obtained with the
aid of the Legendre transforms of $\ln G$ with respect to sources. To obtain rate equations, it is convenient to construct
the generating functional (only the master-equation problem will be discussed here)
with sources linear in $a^+$ and $a$:
\begin{equation}
G(J,J^+) = \!\iint\!{\mathcal D}a^+{\mathcal
D}a\,\,
e^{S(a^+,a)+\int\! (aJ+a^+J^+)dt}\sum_{n=0}^\infty P(0,n)\left[a^+(0)+1\right]^n.\nonumber
\end{equation}
Here, $S(a^+,a)$ is the {\em dynamic action}
determined by the differential operator giving rise to the propagator and the
interaction functional $L_I'$:
\beq
S(a^+,a)=
\int\!d t\left\{\, a^+ \left[-\partial_t-\beta\right]a
+L_I'(a^+,a)\right\}\,.\nonumber
\eeq
The functional Legendre transform is defined  via equations
\begin{equation}
\Gamma(a^+,a)=\ln G(J^+,J)-a^+
J^+-aJ\,,\quad a=\fundoo{\ln G}{J}\,,\quad a^+=\fundoo{\ln G}{J^+}\,,\nonumber
\end{equation}
where the (iterative) solution of the latter with the subsequent substitution in the
expression for $\Gamma(a^+,a)$ is implied.
Seek now $a^+$ and $a$ as the extrema of the functional
$$
\digamma(a^+,a)=\Gamma(a^+,a)+a^+ J^++aJ\,.
$$
This approach has the
advantages that the problem is formulated in terms of explicit functions of the first moments $a^+$ and $a$ of the
random fields and it allows to find possible nonperturbative solutions.

Unfortunately, no closed form is known for the functional $\Gamma(a^+,a)$. The rules for the construction of the perturbative series are,
however, well known \cite{Vasilev98}. The corresponding expression for the functional $\digamma(a^+,a)$ starts
with terms corresponding to the dynamic action and the initial condition, the
rest consisting of an infinite sum of Feynman graphs corresponding to the terms of perturbation expansion:
\beq
\digamma(a^+,a)=S(a^+,a)+a^+
J^++aJ
+\ln\left\{\sum_{n=0}^\infty P(0,n)\left[a^+(0)+1\right]^n\right\}+\,\text{loops}\,.\nonumber
\eeq
For instance,
for the reaction $A+A\to \varnothing$ with the initial Poisson distribution in the
second order in the coupling constant $\lambda$ the {\em effective action} is
\begin{multline}
\Gamma=-\int_0^\infty\!\! dt\!\int\! d{\bf x} \,\left\{a^+\partial_t a
-Da^+\nabla^2a
+\lambda D\left[2a^++
(a^+)^2\right]a^2
\right\}\\
+n_0\int\! d{\bf x}\, a^+({\bf x},0) +{1\over 4} \raisebox{-3.4ex}{
\epsfysize=1.3truecm
\epsffile{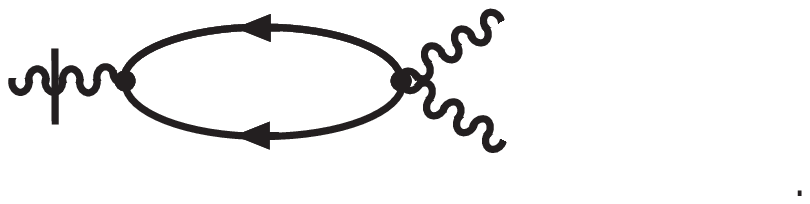}}\negthickspace\negthickspace
\negthickspace\negthickspace\negthickspace\negthickspace\negthickspace\negthickspace
+{1\over 8} \raisebox{-3.4ex}{ \epsfysize=1.3truecm
\epsffile{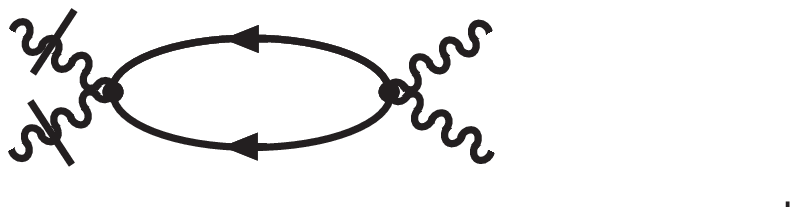}}
\negthickspace\negthickspace\negthickspace\negthickspace\negthickspace\negthickspace
+\ldots\,,\nonumber
\end{multline}
where the wavy line denotes $a$ and the wavy line with slash $a^+$. As a result from variations
with respect to $a$ and $a^+$,
the fluctuation-amended rate equation follows in the form
\beq
\label{RateEqAA1loop}
\partial_t a= D\nabla^2a-2\lambda Da^2
+
4\lambda^2D^2\int_0^\infty\!du\int\!d{\bf y}\,\Delta^2(t-u,\vx-\vy)a^2(u,\vy)+
\ldots,
\eeq
where $\Delta$ is the diffusion kernel $(\partial_t-D\nabla^2)\Delta=1$. The analysis of
the fluctuation term shows that (\ref{RateEqAA1loop}) is not well defined as it stands.
The fluctuation-amended rate equation should be written in terms of a {\em renormalized field theory}.
In that case the asymptotic behaviour of its solution may be analyzed with the aid of the
field-theoretic renormalization group, which has been used for the analysis of the effect of density and drift
fluctuations on the long-time decay already at the rate-equation level \cite{Lee94,Hnatich00}.
The analysis of the renormalized counterpart of (\ref{RateEqAA1loop}) is in progress.

\section{Conclusion}
\label{sec:Conclusion}

Construction of functional representations of solutions of stochastic problems is reviewed.
A unified generic representation of the generating function of Green functions
of random fields brought about either by the
Fokker-Planck equation (corresponding to a Langevin equation) or by the master equation is derived within the
operator approach. Relation between the ambiguities due to the stochastic differential equation and the functional
representation are discussed.
Practically important cases of the Cauchy problem and the scattering problem are presented.
Significance of the choice of the functional representation for numerical calculation of
response and correlation functions is analyzed.
Importance of the variational approach to evaluation of moments of random fields is emphasized.
The fluctuation amended rate equation for the reaction $A+A\to \varnothing$ is discussed.

\subsubsection*{Acknowledgments.} This work was supported by the Finnish Society of Sciences and Letters
(grant number fy2011n24).

\end{document}